\documentclass[11pt,journal,draftcls,onecolumn,peerreviewca]{IEEEtran}

\usepackage{subfigure}
\usepackage{amsfonts}
\usepackage{amsmath}
\usepackage{amssymb}
\usepackage{algorithmic}
\usepackage{algorithm}
\usepackage{graphicx}
\usepackage{color, soul}
\usepackage{stfloats}
\usepackage{setspace}
\usepackage{algorithm}
\usepackage{algorithmic}
\usepackage{float}
\usepackage{verbatim}
\usepackage{cite}
\usepackage{url}
\usepackage{bm}
\usepackage{stmaryrd}
\usepackage{multicol}
\usepackage{stfloats}
\usepackage[amsmath,thmmarks]{ntheorem}
\usepackage{theorem}
\usepackage[justification=centering]{caption}

\theoremheaderfont{\sc}\theorembodyfont{\upshape}
\theoremstyle{nonumberplain}
\theoremseparator{}
\theoremsymbol{\rule{1ex}{1ex}}

\renewcommand{\Re}[1]{\mathrm{Re}\left\{#1\right\}}
\renewcommand{\Im}[1]{\mathrm{Im}\left\{#1\right\}}

\begin{document}

\title{Matrix Completion from Quantized Samples via Generalized Sparse Bayesian Learning}

\author{Jiang Zhu, Zhennan Liu, Qi Zhang, Chunyi Song and Zhiwei Xu  \thanks{The authors are with the Ocean College, Zhejiang University, No.1 Zheda Road, Zhoushan, China, 316021 (e-mail:\{jiangzhu16, lzn2015, zhangqi13, cysong, xuzw\}@zju.edu.cn).
}}

\date{}
\maketitle
\begin{abstract}
The recovery of a low rank matrix from a subset of noisy low-precision quantized samples arises in several applications such as collaborative filtering, intelligent recommendation and millimeter wave channel estimation with few bit ADCs. In this paper, a generalized sparse Bayesian learning (Gr-SBL) combining expectation propagation (EP) is proposed to solve the matrix completion (MC), which is termed as MC-Gr-SBL. The MC-Gr-SBL automatically estimates the rank, the factors and their covariance matrices, and the noise variance. In addition, MC-Gr-SBL is proposed to solve the two dimensional line spectral estimation problem by incorporating the MUSIC algorithm. Finally, substantial numerical experiments are conducted to verify the effectiveness of the proposed algorithm.
\end{abstract}
%
{\bf keywords}: Matrix completion, sparse Bayesian learning, quantization, expectation propagation, two dimensional line spectral estimation
\section{Introduction}
Matrix completion aiming to recover a low-rank matrix from a subset of its entries is a fundamental problem in signal processing and machine learning \cite{RechtMC, ShenMC, MontanariMC, MCfang1}. This problem arises in many applications including sensor networks localization \cite{SNL}, system identification \cite{SI}, collaborative filtering \cite{CF}, millimeter wave channel estimation \cite{MCfang2, MCfang3}. In many of these applications, the entries of the low-rank matrix are not continuous-valued, but discrete or quantized, e.g., binary-valued or multiple-valued. For example, in the Netflix problem where the ratings from the users take integer values between $1$ and $5$. Classical matrix completion treating the values as continuous-valued yields good results \cite{VSBL}, however, performance improvement can be achieved when the observations are treated as quantized \cite{PQMC, RQMC, CMC, RQD, RQEM}.

In \cite{MD1bit}, matrix completion from binary quantized observations is studied. It is shown that noise has a ``dithering'' effect and the problem becomes well-posed. In addition, theoretical analysis reveals that the same degree of accuracy when given access to completely unquantized measurements can be achieved. Later, the extension to multi-bit quantized observations is introduced \cite{RQMC}, where a constrained maximum likelihood estimator similar to \cite{MD1bit} is proposed and its effectiveness is demonstrated numerically. In \cite{PQMC}, theoretical analysis is performed for noisy multi-bit quantization.

Another line of work in the context of both millimeter wave channel estimation and two dimensional line spectral estimation has been concerned with matrix completion. As bandwidths scale up, high resolution ADCs is difficult to implement under cost and power budge constraints. A possible approach is to adopt low resolution ADCs \cite{Singh1, LTL1, Mo}. Since low resolution quantization involves nonlinear operation, conventional matrix completion approach incurs performance loss and designing matrix completion approach which takes quantization effects into consideration will achieve better performance \cite{Bayesjoint}.

In \cite{VSBL}, a novel recovery algorithm termed as variational sparse Bayesian learning (VSBL) for estimating low-rank matrices in matrix completion is proposed. It is numerically shown the effectiveness of the proposed approach in determining the correct rank while providing high recovery performance. Compared to \cite{VSBL}, this work extends the idea and propose a generalized SBL algorithm for matrix completion from quantized observations, termed as MC-Gr-SBL. The MC-Gr-SBL is motivated by the unified inference framework proposed in \cite{UnifedFramework}, which shows that the nonlinear measurement model can be iteratively approximated as a sequence of linear measurement models. In this paper, the nonlinear (quantized) measurement model is iteratively approximated as a sequence of matrix completion from linear measurements with noise being heteroscedastic, i.e., different components having different variance. Therefore, the VSBL is rederived. Since the two dimensional line spectral estimation problem from incomplete measurements can also be viewed as a low rank matrix estimation, MC-Gr-SBL can be applied to estimate the respective subspace. Consequently, multiple signal classification (MUSIC) can then be applied to estimate the line spectral. In addition, a heuristic approach is proposed to obtain the correspondence between the frequencies. The above approach is termed as MC-Gr-SBL-MUSIC. Finally, substantial numerical experiments are conducted to demonstrate the effectiveness of MC-Gr-SBL and MC-Gr-SBL-MUSIC.
\section{Bayesian Modeling for Matrix Completion from Quantized Measurements}
In this section, a hierarchical Bayesian framework is proposed, and the maximum likelihood approach is introduced.
\subsection{Low Rank Modeling}
The low rank parametrization of the unknown matrix ${\mathbf Z}$ is\footnote{Here we study the complex-valued matrix case.}
\begin{align}\label{ZUV}
{\mathbf Z}={\mathbf U}{\mathbf V}^{\rm H},
\end{align}
where ${\mathbf U}$ is an $m\times r$ matrix, and $\mathbf V$ an $n\times r$ matrix such that ${\rm rank}({\mathbf Z})=r\leq {\rm min}(m,n)$. Note that the rank $r$ is unknown. It is clear from ${\mathbf Z}={\mathbf U}{\mathbf V}^{\rm H}$ (\ref{ZUV}) that ${\mathbf Z}$ is the sum of outer-products of the columns of ${\mathbf U}$ and ${\mathbf V}$, i.e., ${\mathbf Z}=\sum\limits_{i=1}^r{\mathbf u}_{\cdot i}{\mathbf v}_{\cdot i}^{\rm H}$. Since in general $r$ is unknown, we introduce an incomplete model and ${\mathbf Z}$ is factorized as
\begin{align}
{\mathbf Z}=\sum\limits_{i=1}^k{\mathbf u}_{\cdot i}{\mathbf v}_{\cdot i}^{\rm H},
\end{align}
where $k$ is known and satisfying $k\geq r$. Similar to the SBL framework, we impose the columns ${\mathbf u}_{\cdot i}$ and ${\mathbf v}_{\cdot i}$ of ${\mathbf U}$ and ${\mathbf V}$ with Gaussian priors of common precisions $\gamma_i$, i.e.,
\begin{align}
p({\mathbf U};{\boldsymbol \gamma})=\prod\limits_{i=1}^k{\mathcal {CN}}({\mathbf u}_{\cdot i}|0,\gamma_i^{-1}{\mathbf I}_m),\\
p({\mathbf V};{\boldsymbol \gamma})=\prod\limits_{i=1}^k{\mathcal {CN}}({\mathbf v}_{\cdot i}|0,\gamma_i^{-1}{\mathbf I}_n).
\end{align}
It can be seen that the columns of $\mathbf U$ and $\mathbf V$ share the same sparsity profile.
\subsection{Observation and Noise Models}
Consider a matrix completion (MC) problem from quantized measurements
\begin{align}\label{model}
{\mathbf Y}_{\Omega}={\mathcal Q}(\Re{{\mathbf Z}_{\Omega}+{\mathbf N}_{\Omega}})+{\rm j}{\mathcal Q}(\Im{{\mathbf Z}_{\Omega}+{\mathbf N}_{\Omega}}),
\end{align}
where ${\Omega}$ is the observed index such that for $(i,j)\in \Omega$, ${Y}_{ij}$ is observed, otherwise $Y_{ij}$ is unobserved, $N_{ij}$ is the independent and identically (i.i.d.) Gaussian noise and satisfies $N_{ij}\sim {\mathcal {CN}}(N_{ij};0,\sigma^2)$, ${\mathcal Q}(\cdot)$ is a quantizer which maps the continuous values into discrete values and the quantization intervals are $\{(t_b,t_{b+1})\}_{b=0}^{|{\mathcal D}|-1}$, where $t_0=-\infty$, $t_{{\mathcal D}}=\infty$, $\bigcup_{b=0}^{{\mathcal D}-1}[t_b,t_{b+1})={\mathbb R}$. The quantized representation of a real number $a\in [t_b,t_{b+1})$ is
\begin{align}
{\mathcal Q}(a)=\omega_b, \quad {\rm if}\quad a\in [t_b,t_{b+1}).
\end{align}
Note that for a quantizer with bit-depth $B$, the cardinality of the output of the quantizer is $|{\mathcal D}|=2^B$. The goal is to reconstruct the low rank matrix ${\mathbf Z}$ through the observed quantized measurements ${\mathbf Y}_{\Omega}$.
\subsection{Maximum Likelihood Estimation}
Given (\ref{model}), the conditional PDF $p({\mathbf Y}|{\mathbf Z};\sigma^2)$ is
\begin{align}
p({\mathbf Y}|{\mathbf Z};\sigma^2)=\prod\limits_{(i,j)\in \Omega}p(Y_{ij}|Z_{ij};\sigma^2),
\end{align}
where $p(Y_{ij}|Z_{ij};\sigma^2)$ can be easily obtained. Let $\boldsymbol \Phi$ and $\boldsymbol \eta$ be
\begin{align}
{\boldsymbol \Phi}=\{{\mathbf Z},{\mathbf U},{\mathbf V}\},\quad {\boldsymbol \eta}=\{{\boldsymbol \gamma},\sigma^2\},
\end{align}
respectively. Therefore, the joint PDF $p({\mathbf Y},{\boldsymbol \Phi};{\boldsymbol \eta})$ is described as
\begin{align}
p({\mathbf Y},{\boldsymbol \Phi};{\boldsymbol \eta})=p({\mathbf Y}|{\mathbf Z};\sigma^2)\delta({\mathbf Z}-{\mathbf U}{\mathbf V}^{\rm H})p({\mathbf U};{\boldsymbol \gamma})p({\mathbf V};{\boldsymbol \gamma}).
\end{align}
In general, the type II maximum likelihood (ML) estimation are adopted to estimate the nuisances parameters ${\boldsymbol \eta}$, i.e.,
\begin{align}\label{MLII}
\hat{\boldsymbol \eta}_{\rm ML}=\underset{{\boldsymbol \eta}}{\operatorname{argmax}}\int p({\mathbf Y},{\boldsymbol \Phi};{\boldsymbol \eta}){\rm d}{\mathbf Z}{\rm d}{\mathbf U}{\rm d}{\mathbf V}.
\end{align}
Then the minimum mean squared error (MMSE) estimate of the parameters ${\boldsymbol \Phi}$ is
\begin{align}\label{MMSEPhi}
(\hat{\mathbf Z},\hat{\mathbf U},\hat{\mathbf V})={\rm E}[({\mathbf Z},{\mathbf U},{\mathbf V})|{\mathbf Y};\hat{\boldsymbol \eta}_{\rm ML}],
\end{align}
where the expectation is taken with respect to
\begin{align}
p({\mathbf Z},{\mathbf U},{\mathbf V}|{\mathbf Y };\hat{\boldsymbol \eta}_{\rm ML})=\frac{p({\mathbf Y},{\mathbf Z},{\mathbf U},{\mathbf V};\hat{\boldsymbol \eta}_{\rm ML})}{p({\mathbf Y };\hat{\boldsymbol \eta}_{\rm ML})}.
\end{align}
However, directly solving the MLE of $\boldsymbol \eta$ (\ref{MLII}) or the MMSE estimate of ${\boldsymbol \Phi}$ (\ref{MMSEPhi}) are both intractable. As a result, approximate Bayesian inference method is adopted.
\section{Inference for MC under Known Heteroscedastic Noise}
In Section \ref{GrMCSBL}, the approximate Bayesian inference approach is derived. The key step is to approximate the quantized (nonlinear) measurement model as a linear measurement model with noise being heteroscedastic. Therefore, this Section studies the MC under known heteroscedastic noise.

For the MC problem, it is described as
\begin{align}\label{pusedoMC}
\widetilde{Y}_{ij}=Z_{ij}+\widetilde{N}_{ij},\cdots, (i,j)\in \Omega,
\end{align}
where $\widetilde{N}_{ij}\sim {\mathcal {CN}}(\widetilde{N}_{ij};0,\beta_{ij}^{-1})$ and $\boldsymbol \beta$ is known. The variational Bayesian inference approach is adopted, and the approximated posterior PDF $q({\mathbf U},{\mathbf V}|\widetilde{\mathbf Y})$ is supposed to be factorized as
\begin{align}
q({\mathbf U},{\mathbf V}|\widetilde{\mathbf Y})=q({\mathbf U}|\widetilde{\mathbf Y})q({\mathbf V}|\widetilde{\mathbf Y}),
\end{align}
i.e., given $\widetilde{\mathbf Y}$, the conditional PDF of ${\mathbf U}$ and ${\mathbf V}$ are independent. According to \cite[pp. 735, eq. (21.25)]{Murphy}, the approximated posterior PDF $q({\mathbf X}|\widetilde{\mathbf Y})$ of each latent variable ${\mathbf X}\in \{{\mathbf U},{\mathbf V}\}$ is
\begin{align}\label{update}
q({\mathbf X}|\widetilde{\mathbf Y})={\rm E}_{q(\{{\mathbf U},{\mathbf V}\}\setminus {\mathbf X}|\widetilde{\mathbf Y})}\left[\log p({\mathbf Y},{\mathbf U},{\mathbf V};{\boldsymbol \eta})\right]+{\rm const},
\end{align}
where $\{{\mathbf U},{\mathbf V}\}\setminus {\mathbf X}$ denotes the set $\{{\mathbf U},{\mathbf V}\}$ with ${\mathbf X}$ removed. Besides, the nuisance parameters $\boldsymbol \gamma$ are estimated via the EM algorithm.
\subsection{Estimation of $\mathbf U$ and $\mathbf V$}
The log likelihood function $\log p(\widetilde{\mathbf Y},{\mathbf U},{\mathbf V};{\boldsymbol \gamma})$ can be written as
\begin{align}\label{likelihood}
\log p(\widetilde{\mathbf Y},{\mathbf U},{\mathbf V};{\boldsymbol \gamma})&=-\sum\limits_{(i,j)\in \Omega}(Y_{ij}-{\mathbf u}_{i\cdot}{\mathbf v_{j\cdot}^{\rm H}})^2\beta_{ij}-\sum\limits_{i=1}^k\gamma_i{\mathbf u}_{\cdot i}^{\rm H}{\mathbf u}_{\cdot i}-\sum\limits_{i=1}^k\gamma_i{\mathbf v}_{\cdot i}^{\rm H}{\mathbf v}_{\cdot i}+{\rm const}\\
&=-\sum\limits_{(i,j)\in \Omega}(Y_{ij}-{\mathbf u}_{i\cdot}{\mathbf v_{j\cdot}^{\rm H}})^2\beta_{ij}-\sum\limits_{l=1}^m{\mathbf u}_{l\cdot}{\boldsymbol \Gamma}{\mathbf u}_{l\cdot}^{\rm H}-\sum\limits_{l=1}^n{\mathbf v}_{l\cdot}{\boldsymbol \Gamma}{\mathbf v}_{l\cdot}^{\rm H}+{\rm const}.
\end{align}
Using (\ref{update}) and (\ref{likelihood}), we update $q({\mathbf u}_{i\cdot })$ as
\begin{align}\label{qui}
q({\mathbf u}_{i\cdot})={\mathcal N}({\mathbf u}_{i\cdot};\hat{\mathbf u}_{i\cdot},{\boldsymbol \Sigma}_i^u),
\end{align}
with mean and covariance as
\begin{subequations}\label{estimateu}
\begin{align}
\hat{\mathbf u}_{i\cdot}&={\boldsymbol \Sigma}_i^u(\sum\limits_{j:(i,j)\in \Omega}\beta_{ij}Y_{ij}\hat{\mathbf v}_{j\cdot})\\
{\boldsymbol \Sigma}_i^u&=(\sum\limits_{j:(i,j)\in \Omega}\beta_{ij}{\rm E}\left[{\mathbf v_{j\cdot}^{\rm H}}{\mathbf v_{j\cdot}}\right]+{\boldsymbol \Gamma})^{-1},
\end{align}
\end{subequations}
where
\begin{align}
{\rm E}\left[{\mathbf v_{j\cdot}^{\rm H}}{\mathbf v_{j\cdot}}\right]=\hat{\mathbf v}_{j\cdot}^{\rm H}\hat{\mathbf v}_{j\cdot}+{\boldsymbol \Sigma}_j^v.
\end{align}
Similarly, we update $q({\mathbf v}_{j\cdot})$ as
\begin{align}\label{qvj}
q({\mathbf v}_{j\cdot})={\mathcal N}({\mathbf v}_{j\cdot};\hat{{\mathbf v}}_{j\cdot},{\boldsymbol \Sigma}_j^v),
\end{align}
with mean and covariance as
\begin{subequations}\label{estimatev}
\begin{align}
\hat{\mathbf v}_{j\cdot}&={\boldsymbol \Sigma}_j^v(\sum\limits_{i:(i,j)\in\Omega}\beta_{ij}Y_{ij}^*\hat{\mathbf u}_{i\cdot})\\
{\boldsymbol \Sigma}_j^v&=(\sum\limits_{i:(i,j)\in\Omega}\beta_{ij}{\rm E}\left[{\mathbf u}_{i\cdot}^{\rm H}{\mathbf u}_{i\cdot}\right]+{\boldsymbol \Gamma})^{-1},
\end{align}
\end{subequations}
where
\begin{align}
{\rm E}\left[{\mathbf u_{i\cdot}^{\rm H}}{\mathbf u_{i\cdot}}\right]=\hat{\mathbf u}_{i\cdot}^{\rm H}\hat{\mathbf u}_{i\cdot}+{\boldsymbol \Sigma}_i^u.
\end{align}
\subsection{Estimation of $\boldsymbol \gamma$}
The EM approach is adopted to estimate $\boldsymbol \gamma$. First, the averaged complete loglikelihood function $\log p(\widetilde{\mathbf Y},{\mathbf U},{\mathbf V};{\boldsymbol \gamma})$ with respect to the PDF $q({\mathbf U}|\widetilde{\mathbf Y};{\boldsymbol \gamma}_{\rm old})q({\mathbf V}|\widetilde{\mathbf Y};{\boldsymbol \gamma}_{\rm old})$ is evaluated as
\begin{align}
Q({\boldsymbol \gamma};{\boldsymbol \gamma}_{\rm old})=-\sum\limits_{i=1}^k\left(\gamma_i{\rm E}\left[{\mathbf u}_{\cdot i}^{\rm H}{\mathbf u}_{\cdot i}\right]+\gamma_i{\rm E}\left[{\mathbf v}_{\cdot i}^{\rm H}{\mathbf v}_{\cdot i}\right]-(m+n)\log \gamma_i\right).
\end{align}
Setting $\partial Q({\boldsymbol \gamma};{\boldsymbol \gamma}_{\rm old})/\partial {\boldsymbol \gamma}={\mathbf 0}$ yields
\begin{align}\label{updategamma}
\gamma_i=\frac{m+n}{{\rm E}\left[{\mathbf u}_{\cdot i}^{\rm H}{\mathbf u}_{\cdot i}\right]+{\rm E}\left[{\mathbf v}_{\cdot i}^{\rm H}{\mathbf v}_{\cdot i}\right]}=\frac{m+n}{\hat{\mathbf u}_{\cdot i}^{\rm H}\hat{\mathbf u}_{\cdot i}+\sum\limits_j ({\boldsymbol \Sigma}_j^u)_{ii}+\hat{\mathbf v}_{\cdot i}^{\rm H}\hat{\mathbf v}_{\cdot i}+\sum\limits_j ({\boldsymbol \Sigma}_j^v)_{ii}}.
\end{align}
In summary, the algorithm proceeds by first estimating the rows of $\mathbf U$ and $\mathbf V$ through (\ref{estimateu}) and (\ref{estimatev}), respectively, and then followed by the updating of precisions $\boldsymbol \gamma$ (\ref{updategamma}).
\section{Gr-SBL-MC from Quantized Samples}\label{GrMCSBL}
\begin{figure}
  \centering
  \includegraphics[width=3.6in]{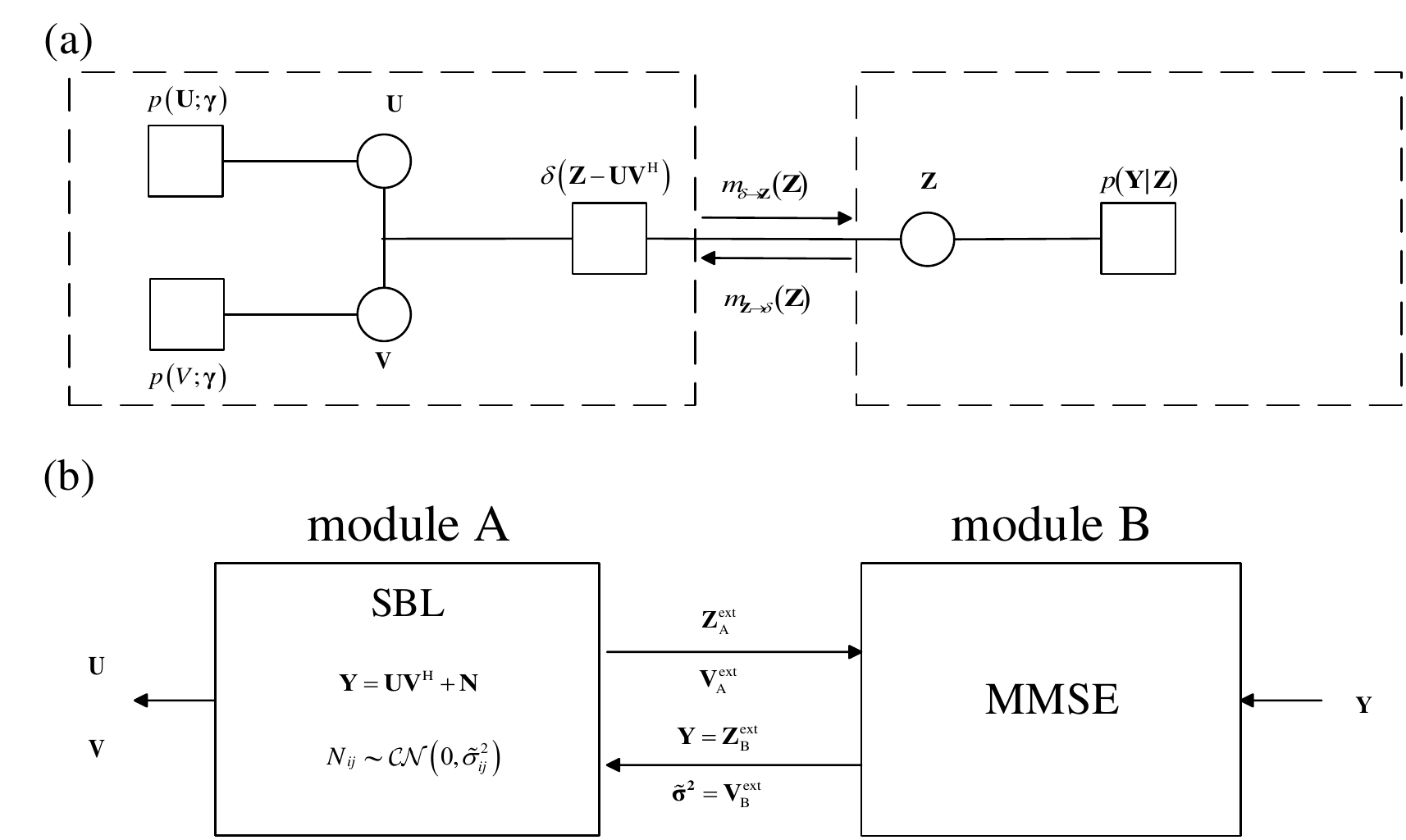}\\
  \caption{The original factor graph for the MC problem}\label{module1}
\end{figure}

As shown in \cite{UnifedFramework}, a unified Bayesian inference framework is proposed to solve the generalized linear model through standard approximate Bayesian inference algorithm. The key idea is to iteratively approximate the quantized model as a standard MC model with heteroscedastic noise (different components having different variance). The factor graph for the proposed Bayesian modeling is presented in Fig. \ref{module1} (a). The algorithm composes of two modules named module A and module B, where module A runs the standard approximate Bayesian algorithm, module B performs the MMSE estimation to refine the pseudo observations and noise variances of the linear model in module A, see Fig. \ref{module1} (b). The two modules exchange their extrinsic information and iterate until convergence or the stopping criterion is satisfied. Here we follow the idea and propose the Gr-SBL-MC for matrix completion from quantized samples.

\subsection{MMSE module}
Let the extrinsic message $m_{\delta\rightarrow {\mathbf Z}}({\mathbf Z})$ transmitted from the factor node $\delta({\mathbf Z}-{\mathbf U}{\mathbf V}^{\rm H})$ to the variable node ${\mathbf Z}$ be $m_{\delta\rightarrow {\mathbf Z}}({\mathbf Z})=\prod\limits_{(i,j)\in \Omega}m_{\delta\rightarrow Z_{ij}}(Z_{ij})$, where
\begin{align}
m_{\delta\rightarrow Z_{ij}}(Z_{ij})={\mathcal {CN}}(Z_{ij};Z_{{\rm A},ij}^{\rm ext},V_{{\rm A},ij}^{\rm ext}),
\end{align}
$Z_{{\rm A},ij}^{\rm ext}$ and $V_{{\rm A},ij}^{\rm ext}$ denote the extrinsic mean and variance of module A, respectively. For the measurement model (\ref{model}), the likelihood function $p({\mathbf Y}|{\mathbf Z})$ is obtained given ${\mathbf Z}$. According to EP, the message $m_{{\mathbf Z}\rightarrow\delta}({\mathbf Z})$ transmitted from the variable node ${\mathbf Z}$ to the factor node $\delta({\mathbf Z}-{\mathbf U}{\mathbf V}^{\rm H})$ is
\begin{align}\label{ext_m_z}
m_{{\mathbf Z}\rightarrow \delta}({\mathbf Z})\propto&\frac{{\rm Proj}\left[m_{\delta\rightarrow {\mathbf Z}}({\mathbf Z}) p({\mathbf Y}|{\mathbf Z})\right]}{m_{\delta\rightarrow {\mathbf Z}}({\mathbf Z})}\propto \frac{{\rm Proj}\left[\prod\limits_{(i,j)\in \Omega}m_{\delta\rightarrow Z_{ij}}(Z_{ij}) p({Y}_{ij}|{Z}_{ij})\right]}{\prod\limits_{(i,j)\in \Omega}m_{\delta\rightarrow Z_{ij}}(Z_{ij})}\notag\\
=&\prod\limits_{(i,j)\in \Omega}\frac{{\rm Proj}\left[m_{\delta\rightarrow Z_{ij}}(Z_{ij}) p({Y}_{ij}|{Z}_{ij})\right]}{m_{\delta\rightarrow Z_{ij}}(Z_{ij})}\triangleq \prod\limits_{(i,j)\in \Omega}\frac{{\rm Proj}\left[q_{\rm B}(Z_{ij})\right]}{m_{\delta\rightarrow Z_{ij}}(Z_{ij})}\notag\\
\propto &\prod\limits_{(i,j)\in \Omega}m_{Z_{ij}\rightarrow \delta}(Z_{ij})\triangleq {\mathcal {CN}}(Z_{ij};Z_{{\rm B},ij}^{\rm ext},V_{{\rm B},ij}^{\rm ext}).
\end{align}
Let $m_{\delta\rightarrow {\mathbf Z}}({\mathbf Z})=\prod\limits_{(i,j)\in \Omega}m_{\delta\rightarrow Z_{ij}}(Z_{ij})$ be the prior of ${\mathbf Z}$. Combining the likelihood $p({\mathbf Y}|{\mathbf Z})=\prod\limits_{(i,j)\in \Omega} p(Y_{ij}|Z_{ij})$, the componentwise MMSE of ${\mathbf Z}$ can be obtained as
\begin{align}
&Z_{{\rm B},ij}^{\rm post}={\rm E}\left[Z_{ij}|q_{\rm B}(Z_{ij})\right],\label{postZB}\\
&V_{{\rm B},ij}^{\rm post}={\rm Var}\left[Z_{ij}|q_{\rm B}(Z_{ij})\right].\label{postVB}
\end{align}
As a result, ${\rm Proj}\left[m_{\delta\rightarrow Z_{ij}}^t(Z_{ij})p({Y}_{ij}|{Z}_{ij})\right]$ is
\begin{align}
{\rm Proj}\left[m_{\delta\rightarrow Z_{ij}}(Z_{ij})p({Y}_{ij}|{Z}_{ij})\right]={\mathcal {CN}}(Z_{ij};Z_{{\rm B},ij}^{\rm post},V_{{\rm B},ij}^{\rm post}).
\end{align}
From (\ref{ext_m_z}), $m_{Z_{ij}\rightarrow \delta}(Z_{ij})$ is calculated as
\begin{align}
&\frac{1}{V_{{\rm B},ij}^{\rm ext}}=\frac{1}{V_{{\rm B},ij}^{\rm post}}-\frac{1}{V_{{\rm A},ij}^{\rm ext}},\label{VBext}\\
&Z_{{\rm B},ij}^{\rm ext}=V_{{\rm B},ij}^{\rm ext}\left(\frac{Z_{{\rm B},ij}^{\rm post}(t)}{V_{{\rm B},ij}^{\rm post}}-\frac{Z_{{\rm A},ij}^{\rm ext}}{V_{{\rm A},ij}^{\rm ext}}\right).\label{ZBext}
\end{align}
\subsection{MC Module}
According to the definition of $\delta({\mathbf Z}-{\mathbf U}{\mathbf V}^{\rm H})$ and $m_{{\mathbf Z}\rightarrow\delta }({\mathbf Z})$ and eliminating the variable node $\mathbf Z$, the pseudo linear measurement model (\ref{pusedoMC}) is obtained, where
\begin{align}
\tilde{\mathbf Y}&={\mathbf Z}_{{\rm B}}^{\rm ext},\\
{\boldsymbol \beta}&=1/{\mathbf V}_{{\rm B}}^{\rm ext}.
\end{align}
After running a single iteration of the variational Bayesian method, the posterior PDF $q({\mathbf U}|\tilde{\mathbf Y})$ and $q({\mathbf V}|\tilde{\mathbf Y})$ are obtained. According to EP, the message $m_{\delta\rightarrow{\mathbf Z}}({\mathbf Z})$ from MC module to MMSE module is calculated as
\begin{align}\label{mextA}
m_{\delta\rightarrow{\mathbf Z}}({\mathbf Z})&=\frac{{\rm Proj}\left[\int m_{{\mathbf Z}\rightarrow \delta}({\mathbf Z})p({\mathbf U};{\boldsymbol \gamma})p({\mathbf V};{\boldsymbol \gamma})\delta({\mathbf Z}-{\mathbf U}{\mathbf V}^{\rm H}){\rm d}{\mathbf U}{\rm d}{\mathbf V}\right]}{m_{{\mathbf Z}\rightarrow \delta}({\mathbf Z})}\notag\\
&=\frac{{\rm Proj}\left[q_{\rm A}({\mathbf Z})\right]}{m_{{\mathbf Z}\rightarrow \delta}({\mathbf Z})}.
\end{align}
The posterior means and variances of $\mathbf Z$ can be calculated as
\begin{align}\label{postZa}
&{\mathbf Z}_{\rm A}^{\rm post}=\hat{\mathbf U}\hat{\mathbf V}^{\rm H},\\
&V_{{\rm A},ij}^{\rm post}={\mathbf v}_{j\cdot}{\boldsymbol \Sigma}_i^u{\mathbf v}_{j\cdot}^{\rm H}+{\mathbf u}_{i\cdot}{\boldsymbol \Sigma}_j^v{\mathbf u}_{i\cdot}^{\rm H}+{\rm tr}\left({\boldsymbol \Sigma}_i^u{\boldsymbol \Sigma}_j^v\right).
\end{align}
Thus
\begin{align}
{\rm Proj}\left[q_{\rm A}({\mathbf Z})\right]=\prod\limits_{i}\prod\limits_{j}{\mathcal {CN}}\left(Z_{ij};Z_{{\rm A},ij}^{\rm post};V_{{\rm A},ij}^{\rm post}\right).
\end{align}
According to (\ref{mextA}), $m_{{\mathbf Z}\rightarrow \delta}({\mathbf Z})=\prod\limits_{i}\prod\limits_{j}{\mathcal {CN}}\left(Z_{ij};Z_{{\rm A},ij}^{\rm ext};V_{{\rm A},ij}^{\rm ext}\right)$, $Z_{{\rm A},ij}^{\rm ext}$ and $V_{{\rm A},ij}^{\rm ext}$ are given by
\begin{align}
&{V_{{\rm A},ij}^{\rm ext}}=\left(\frac{1}{V_{{\rm A},ij}^{\rm post}}-\frac{1}{V_{{\rm B},ij}^{\rm ext}}\right)^{-1},\label{extvarzA}\\
&Z_{{\rm A},ij}^{\rm ext}=V_{{\rm A},ij}^{\rm ext}\left(\frac{Z_{{\rm A},ij}^{\rm post}}{V_{{\rm A},ij}^{\rm post}}-\frac{Z_{{\rm B},ij}^{\rm ext}}{V_{{\rm B},ij}^{\rm ext}}\right),\label{extmeanzA}
\end{align}
which closes the loop of the proposed Gr-SBL-MC algorithm. In addition, EM algorithm can also be adopted to learn the noise variance $\sigma^2$ as\footnote{For one-bit quantization, numerical results show that it is better to perform the MC-Gr-SBL without estimating the noise variance $\sigma^2$.}
\begin{align}\label{sigmahat}
\hat{\sigma}^2=\frac{\sum\limits_{(i,j)\in \Omega}\left(|Z_{{\rm B},ij}^{\rm ext}-{Z}_{{\rm A},ij}^{\rm post}|^2+V_{{\rm A},ij}^{\rm post}\right)}{|{\Omega}|}.
\end{align}
\begin{algorithm}[h]
\caption{MC-Gr-SBL algorithm}
\begin{algorithmic}[1]
\STATE Initialize ${\mathbf V}_{\rm A}^{\rm ext}$, ${\mathbf Z}_{\rm A}^{\rm ext}$, $\hat{\sigma}^2$, then perform the MMSE estimation and obtain the post mean and variance of $\mathbf Z$ as ${\mathbf V}_{\rm B}^{\rm post}$ (\ref{postVB}) and ${\mathbf Z}_{\rm B}^{\rm post}$ (\ref{postZB}), finally compute the extrinsic mean and variance of $\mathbf Z$ as ${\mathbf V}_{\rm B}^{\rm ext}$ (\ref{VBext}), ${\mathbf Z}_{\rm B}^{\rm ext}(t)$ (\ref{ZBext}). \
\FOR {$t=1,\cdots,T_{\rm outer}$ }
\STATE Update $q({\mathbf u}_{i\cdot})$ (\ref{qui}), $q({\mathbf v}_{j\cdot})$ (\ref{qvj}), $\hat{\boldsymbol \gamma}$ (\ref{updategamma}).
\STATE Compute the post mean and variance of $\mathbf Z$ as ${\mathbf V}_{\rm A}^{\rm post}$ and ${\mathbf Z}_{\rm A}^{\rm post}$ (\ref{postZa}). Besides, update the noise variance $\hat{\sigma}^2$ as (\ref{sigmahat}).
\STATE Compute the extrinsic mean and variance of $\mathbf Z$ as ${\mathbf V}_{\rm A}^{\rm ext}$ (\ref{extvarzA}), ${\mathbf Z}_{\rm A}^{\rm ext}$ (\ref{extmeanzA}).
\STATE Perform the MMSE estimation and obtain the post mean and variance of $\mathbf Z$ as ${\mathbf V}_{\rm B}^{\rm post}$ (\ref{postVB}) and ${\mathbf Z}_{\rm B}^{\rm post}$ (\ref{postZB})
\STATE Compute the extrinsic mean and variance of $\mathbf Z$ as ${\mathbf V}_{\rm B}^{\rm ext}$ (\ref{VBext}), ${\mathbf Z}_{\rm B}^{\rm ext}(t)$ (\ref{ZBext}).\
\ENDFOR
\STATE Return $\hat{\mathbf Z}$ and ${\rm rank}(\hat{\mathbf Z})$.
\end{algorithmic}
\end{algorithm}
\section{Extension to the 2D Line Spectral Estimation}
For the 2D line spectral estimation problem, the line spectral is
\begin{align}\label{2DLSE}
{\mathbf Z}=\sum\limits_{i=1}^r g_i{\mathbf a}_m(\theta_i){\mathbf a}_n^{\rm H}(\phi_i)={\mathbf A}_m(\boldsymbol \theta){\rm diag}({\mathbf g}){\mathbf A}_n^{\rm H}(\boldsymbol \phi),
\end{align}
where ${\mathbf a}_m(\theta)=[1,{\rm e}^{{\rm j}\theta},\cdots,{\rm e}^{{\rm j}(m-1)\theta}]^{\rm T}$. In general $r\ll {\min}(m,n)$, thus the line spectral $\mathbf Z$ can be viewed as a low rank matrix. Suppose that the measurement model is (\ref{model}), i.e., incomplete quantized measurements are obtained. In practice, this may correspond to a scenario where sparse planar array is employed.

Since $\mathbf Z$ can be described as ${\mathbf Z}={\mathbf U}{\mathbf V}^{\rm H}$ (\ref{ZUV}), the Gr-SBL-MC approach can be used to obtain $\hat{\mathbf U}$ and $\hat{\mathbf V}$. Compared to (\ref{2DLSE}), it is concluded that there exists a unitary matrix ${\boldsymbol \Gamma}$ satisfying ${\boldsymbol \Gamma}{\boldsymbol \Gamma}^{\rm H}={\boldsymbol \Gamma}^{\rm H}{\boldsymbol \Gamma}={\mathbf I}$ such that
\begin{align}
\hat{\mathbf U}&\approx{\mathbf A}_m(\hat{\boldsymbol \theta}){\rm diag}(\hat{\mathbf f}){\boldsymbol \Gamma},\label{Un}\\
\hat{\mathbf V}&\approx{\mathbf A}_n(\hat{\boldsymbol \phi}){\rm diag}(|\hat{\mathbf h}|){\rm diag}({\rm e}^{{\rm j}\angle{\hat{\mathbf h}}}){\boldsymbol \Gamma},\label{Vn}
\end{align}
where $\hat{\mathbf f}\in {\mathbb R}^r$, $\hat{\mathbf h}\in {\mathbb C}^r$ and $\hat{\mathbf g}\approx\hat{\mathbf f}|\hat{\mathbf h}|{\rm e}^{-{\rm j}\angle{\hat{\mathbf h}}}$. Obviously, $\hat{\mathbf U}\hat{\mathbf V}^{\rm H}\approx{\mathbf A}_m(\hat{\boldsymbol \theta}){\rm diag}(\hat{\mathbf g}){\mathbf A}_n^{\rm H}(\hat{\boldsymbol \phi})$.

In the following, the Gr-SBL-MC is proposed to solve the 2D line spectral estimation problem, i.e., recovering $\{(\theta_i,\phi_i)\}_{i=1}^r$ and $r$. Obviously, $\hat{r}$ equals to the rank of $\hat{\mathbf Z}$. Also,
\begin{align}\label{UVvector}
\hat{\mathbf U}\hat{\mathbf U}^{\rm H}&\approx\hat{\mathbf A}_m(\boldsymbol \theta){\rm diag}(|\hat{\mathbf f}|^2)\hat{\mathbf A}_m^{\rm H}(\boldsymbol \theta)=\sum\limits_{i=1}^r\hat{f}_i^2{\mathbf a}_m(\hat{\theta}_i){\mathbf a}_m^{\rm H}(\hat{\theta}_i),\\
\hat{\mathbf V}\hat{\mathbf V}^{\rm H}&\approx\hat{\mathbf A}_n(\hat{\boldsymbol \phi}){\rm diag}(|\hat{\mathbf h}|^2)\hat{\mathbf A}_n^{\rm H}(\hat{\boldsymbol \phi})=\sum\limits_{i=1}^r|\hat{h}_i|^2{\mathbf a}_n(\hat{\phi}_i){\mathbf a}_n^{\rm H}(\hat{\phi}_i).
\end{align}
Thus we use MUSIC to obtain $\hat{\boldsymbol \theta}$ and $\hat{\boldsymbol \phi}$. By vectorizing (\ref{UVvector}), we have
\begin{align}
{\rm vec}(\hat{\mathbf U}\hat{\mathbf U}^{\rm H})\approx[{\mathbf a}_m^*(\hat{\theta}_1)\otimes{\mathbf a}_m(\hat{\theta}_1),\cdots,{\mathbf a}_m^*(\hat{\theta}_i)\otimes{\mathbf a}_m(\hat{\theta}_i),\cdots,{\mathbf a}_m^*(\hat{\theta}_r)\otimes{\mathbf a}_m(\hat{\theta}_r)]\hat{\mathbf f}^2,\notag\\
{\rm vec}(\hat{\mathbf V}\hat{\mathbf V}^{\rm H})\approx[{\mathbf a}_n^*(\hat{\phi}_1)\otimes{\mathbf a}_n(\hat{\phi}_1),\cdots,{\mathbf a}_n^*(\hat{\phi}_i)\otimes{\mathbf a}_n(\hat{\phi}_i),\cdots,{\mathbf a}_n^*(\hat{\phi}_r)\otimes{\mathbf a}_n(\hat{\phi}_r)]|\hat{\mathbf h}|^2
\end{align}
Then LS approach can be used to obtain $\hat{\mathbf f}$ and $|\hat{\mathbf h}|^2$. Since $\boldsymbol \theta$ and $\boldsymbol \phi$ are estimated independently, the correspondence between $\hat{\boldsymbol \theta}$ and $\hat{\boldsymbol \phi}$ is unknown. Thus, an unknown permutation matrix where all its components are either $0$ or $1$ and each row and each column has exactly one nonzero element is introduced to describe the correspondence.

Now we estimate the permutation matrix. Firstly, the unitary matrix ${\boldsymbol \Gamma}$ is estimated as
\begin{align}
\hat{\boldsymbol \Gamma}=\left({\mathbf A}_m(\hat{\boldsymbol \theta}){\rm diag}(\hat{\mathbf f})\right)^{\dagger}\hat{\mathbf U}
\end{align}
according to (\ref{Un}). Then, according to (\ref{Vn}), given $\hat{\boldsymbol \theta}$, $\hat{\boldsymbol \phi}$, $\hat{\mathbf f}$, $|\hat{\mathbf h}|$, $\hat{\boldsymbol \Gamma}$, we hope to find a permutation matrix $\boldsymbol \Pi$ such that
\begin{align}
\hat{\mathbf V}\approx \hat{\mathbf A}_n(\hat{\boldsymbol \phi}){\rm diag}(|\hat{\mathbf h}|){\rm diag}({\rm e}^{{\rm j}\angle{{\mathbf h}}}){\boldsymbol \Pi}\hat{\boldsymbol \Gamma}.\label{VnPi}
\end{align}
or
\begin{align}
\hat{\mathbf V}{\boldsymbol \Gamma}^{\rm H}\approx \hat{\mathbf A}_n(\hat{\boldsymbol \phi}){\rm diag}(|\hat{\mathbf h}|){\rm diag}({\rm e}^{{\rm j}\angle{{\mathbf h}}})\hat{\boldsymbol \Pi}.\label{VnPiv2}
\end{align}
Since ${\boldsymbol \Pi}{\boldsymbol \Pi}^{\rm T}={\mathbf I}$, we obtain
\begin{align}
\hat{\mathbf V}\hat{\boldsymbol \Gamma}^{\rm H}\hat{\boldsymbol \Gamma}^{*}\hat{\mathbf V}^{\rm T}\approx {\mathbf A}_n(\hat{\boldsymbol \phi}){\rm diag}(|\hat{\mathbf h}|^2){\rm diag}({\rm e}^{{\rm j}2\angle{\hat{\mathbf h}}}){\mathbf A}_n^{\rm T}(\hat{\boldsymbol \phi}).\label{VnPiElimPi}
\end{align}
Through vectorizing (\ref{VnPiElimPi}), we have
\begin{align}
{\rm vec}(\hat{\mathbf V}{\boldsymbol \Gamma}^{\rm H}{\boldsymbol \Gamma}^{*}\hat{\mathbf V}^{\rm T})\approx[|\hat{h}_1|^2{\mathbf a}_n(\hat{\phi}_1)\otimes{\mathbf a}_n(\hat{\phi}_1),\cdots,|\hat{h}_i|^2{\mathbf a}_n(\hat{\phi}_i)\otimes{\mathbf a}_n(\hat{\phi}_i),\cdots,|\hat{h}_r|^2{\mathbf a}_n(\hat{\phi}_r)\otimes{\mathbf a}_n(\hat{\phi}_r)]{\rm e}^{{\rm j}2\angle{{\mathbf h}}}.
\end{align}
Using LS we obtain ${\rm e}^{{\rm j}2\angle{\hat{\mathbf h}}}$, where $2\angle{\hat{\mathbf h}}\in [-\pi,\pi)$. Since $\angle{{\mathbf h}}\in [-\pi,\pi)$, it can be easily shown that ${\rm e}^{{\rm j}\angle{{\mathbf h}}}$ equals to either $\sqrt{{\rm e}^{{\rm j}2\angle{\hat{\mathbf h}}}}$ or $-\sqrt{{\rm e}^{{\rm j}2\angle{\hat{\mathbf h}}}}$, depending on $\angle{{\mathbf h}}\in [-\pi/2,\pi/2]$ or $\angle{{\mathbf h}}\in [-\pi,-\pi/2]\cup[\pi/2,\pi]$. Thus
\begin{align}
{\rm e}^{{\rm j}\angle{{\mathbf h}}}={\mathbf J}\sqrt{{\rm e}^{{\rm j}2\angle{\hat{\mathbf h}}}},
\end{align}
where $\mathbf J$ is a diagonal matrix with elements being either $1$ or $-1$. According to (\ref{VnPiv2}), we have
\begin{align}
\hat{\mathbf V}\hat{\boldsymbol \Gamma}^{\rm H}\approx {\mathbf A}_n(\hat{\boldsymbol \phi}){\rm diag}(|\hat{\mathbf h}|){\rm diag}(\sqrt{{\rm e}^{{\rm j}2\angle{\hat{\mathbf h}}}}){\mathbf J}{\boldsymbol \Pi}\triangleq {\mathbf A}_n(\hat{\boldsymbol \phi}){\rm diag}(|\hat{\mathbf h}|){\rm diag}(\sqrt{{\rm e}^{{\rm j}2\angle{\hat{\mathbf h}}}}){\mathbf J}_{\Pi},\label{VnPiv3}
\end{align}
where ${\mathbf J}_{\Pi}$ is a generalized permutation matrix such that all its components are either $0$, $1$, or $-1$ and each row and each column has exactly one nonzero element \cite[Definition 7.6]{AmirBook}. Using LS we obtain $\hat{\mathbf J}_{\Pi}$. Then we use an heuristic approach to approximate $\hat{\mathbf J}_{\Pi}$ which proceeds as follows: Firstly, finding the maximum amplitude of the whole elements, record its position and return its sign by comparing the distance of $-1$ and $1$. Then delete the selected row and column and continue the step. Finally, we obtain $\hat{\mathbf J}_{\Pi}$, the correspondence between $\hat{\boldsymbol \theta}$ and $\hat{\boldsymbol \phi}$ and the phase of $\mathbf h$ are obtained. Finally, the 2D line spectral estimation problem is solved.

\section{Numerical Simulation}
In this section, substantial numerical experiments are conducted to investigate the performance of the proposed algorithm. As for the quantizer, zero threshold is chosen for 1-bit quantizer, while a uniform quantizer is chosen for multi-bit quantization. Let $\sigma_z^2$ be the variance of the elements of $\mathbf Z$. Since the real and imaginary parts of $\mathbf Z$ are quantized separately, and the dynamic range of the quantizer is restricted to be $[-3\sigma_z/\sqrt{2},3\sigma_z/\sqrt{2}]$. For a uniform quantizer with bit-depth $B$, the quantizer step size $\Delta$ is $\Delta=3\sigma_z/2^{B-0.5}$.
Note that for randomly generated matrices ${\mathbf Z}={\mathbf U}{\mathbf V}^{\rm H}$ where the elements of ${\mathbf U}$ and ${\mathbf V}$ are drawn i.i.d. from ${\mathcal {CN}}(0,1)$, then straightforward calculation shows that $\sigma_z^2=r$. For the 2D line spectral signal (\ref{2DLSE}) where the elements of the magnitude $|\mathbf g|$ are drawn from ${\mathcal {CN}}(1,0.2)$ and the phase $\angle\mathbf g$ are drawn uniformly from $[-\pi,\pi]$, it can be shown that $\sigma_z^2\approx r$.
%

For randomly generated matrices, the normalized mean-squared error (NMSE) ${\rm NMSE}(\hat{\mathbf Z})=20\log\frac{\|\hat{\mathbf Z}-{\mathbf Z}\|_{\rm F}}{\|{\mathbf Z}\|_{\rm F}}$, the correct rank estimation probability ${\rm P}(\hat{r}=r)$. Please note that, due to magnitude ambiguity, it is impossible to recover the exact magnitude of the matrix $\mathbf Z$ from one-bit measurements in the noiseless case. Thus for one bit quantization, the debiased NMSE of the matrix is defined as $\min \limits_c 10\log({\|\mathbf Z-c\hat{\mathbf Z}\|_2}/{\|\mathbf Z\|_2})$. For the real matrices set, the NMAE and RMSE are used. For the 2D LSE problem, the normalized mean-squared error(NMSE) ${\rm NMSE}(\hat{\mathbf Z})=20\log\frac{\|\hat{\mathbf Z}-{\mathbf Z}\|_{\rm F}}{\|{\mathbf Z}\|_{\rm F}}$ and the correct rank estimation probability ${\rm P}(\hat{r}=r)$ are adopted as performance metrics. In addition, the MSEs of $\hat{\boldsymbol \theta}$ and $\hat{\boldsymbol \phi}$ defined as ${\rm MSE}(\hat{\boldsymbol \theta})=10\log\|\hat{\boldsymbol \theta}-{\boldsymbol \theta}\|^2$ and ${\rm MSE}(\hat{\boldsymbol \phi})=10\log\|\hat{\boldsymbol \phi}-{\boldsymbol \phi}\|^2$ averaged over the trials in which $\hat{r}=r$ are adopted as performance metrics. All the results are averaged over $50$ Monte Carlo (MC) trials unless stated otherwise.
\subsection{Evaluation Under Random Measurement Matrix}
This subsection investigates the performance of MC-Gr-SBL versus the factors such as bit depth, SNR, fraction of entries sampled $p=\Omega/(mn)$, the number of rows $m$, the rank $r$.

\begin{figure}[htbp]
\centering
\subfigure{
\includegraphics[width=1.8in]{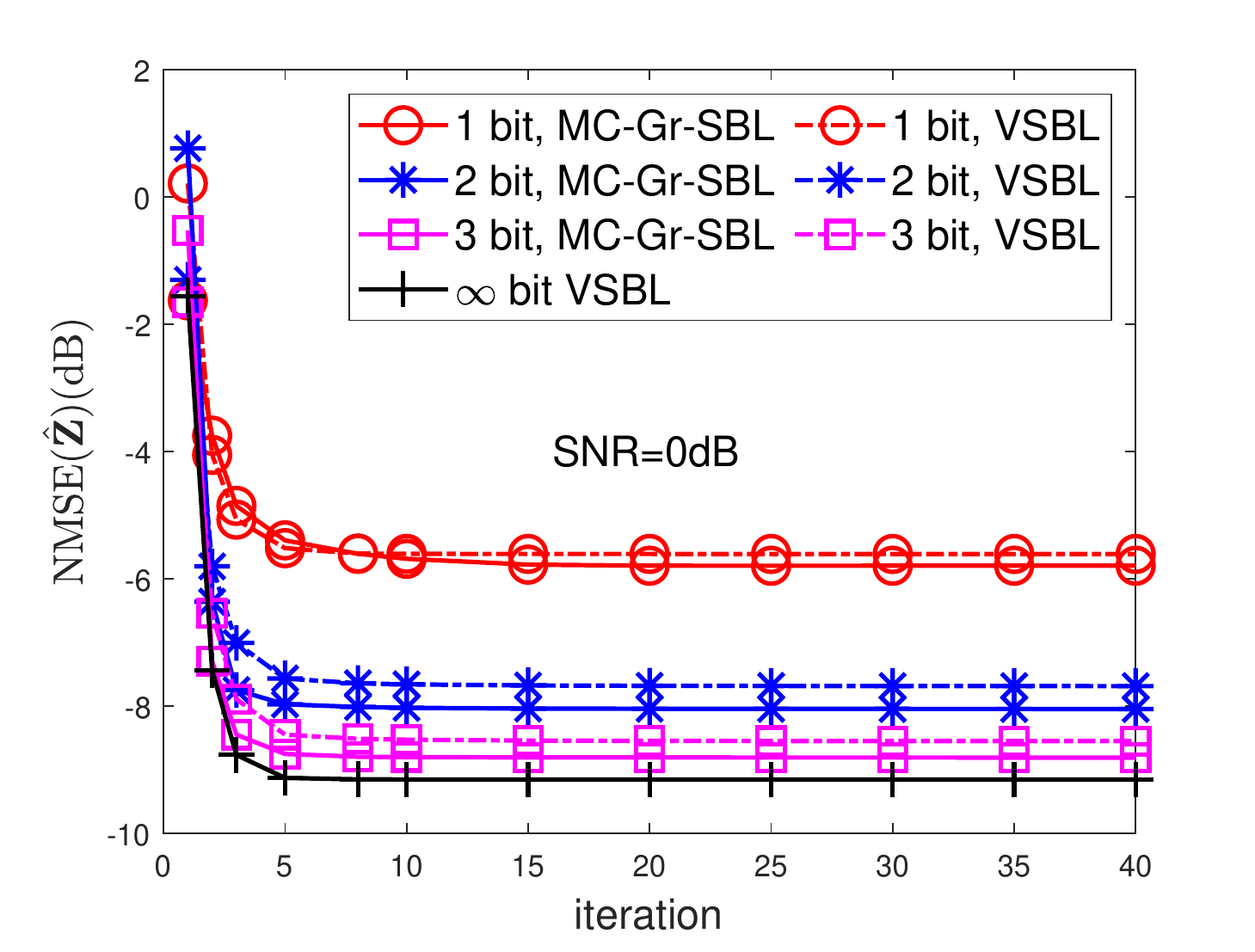}
}
\quad
\subfigure{
\includegraphics[width=1.8in]{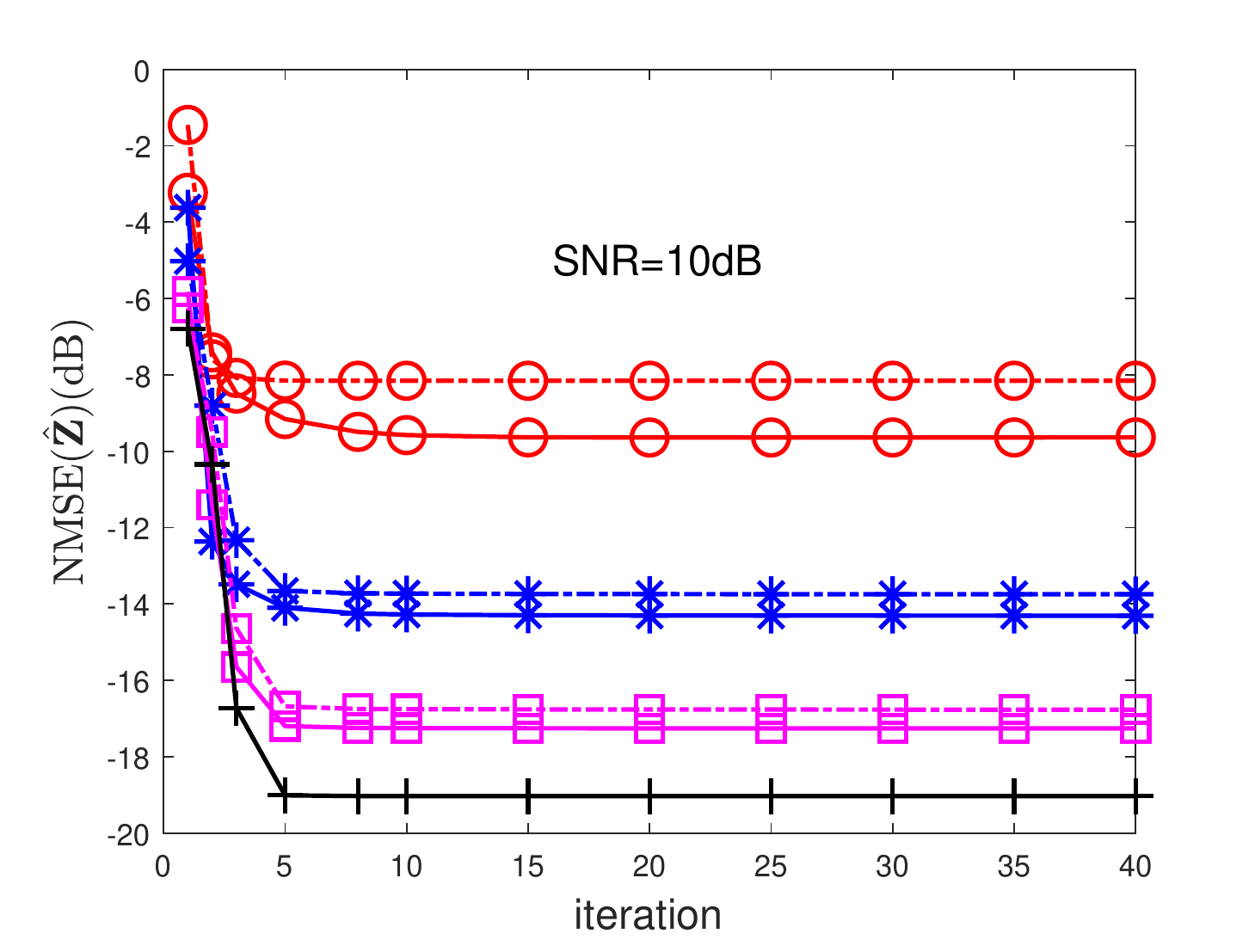}
}
\quad
\subfigure{
\includegraphics[width=1.8in]{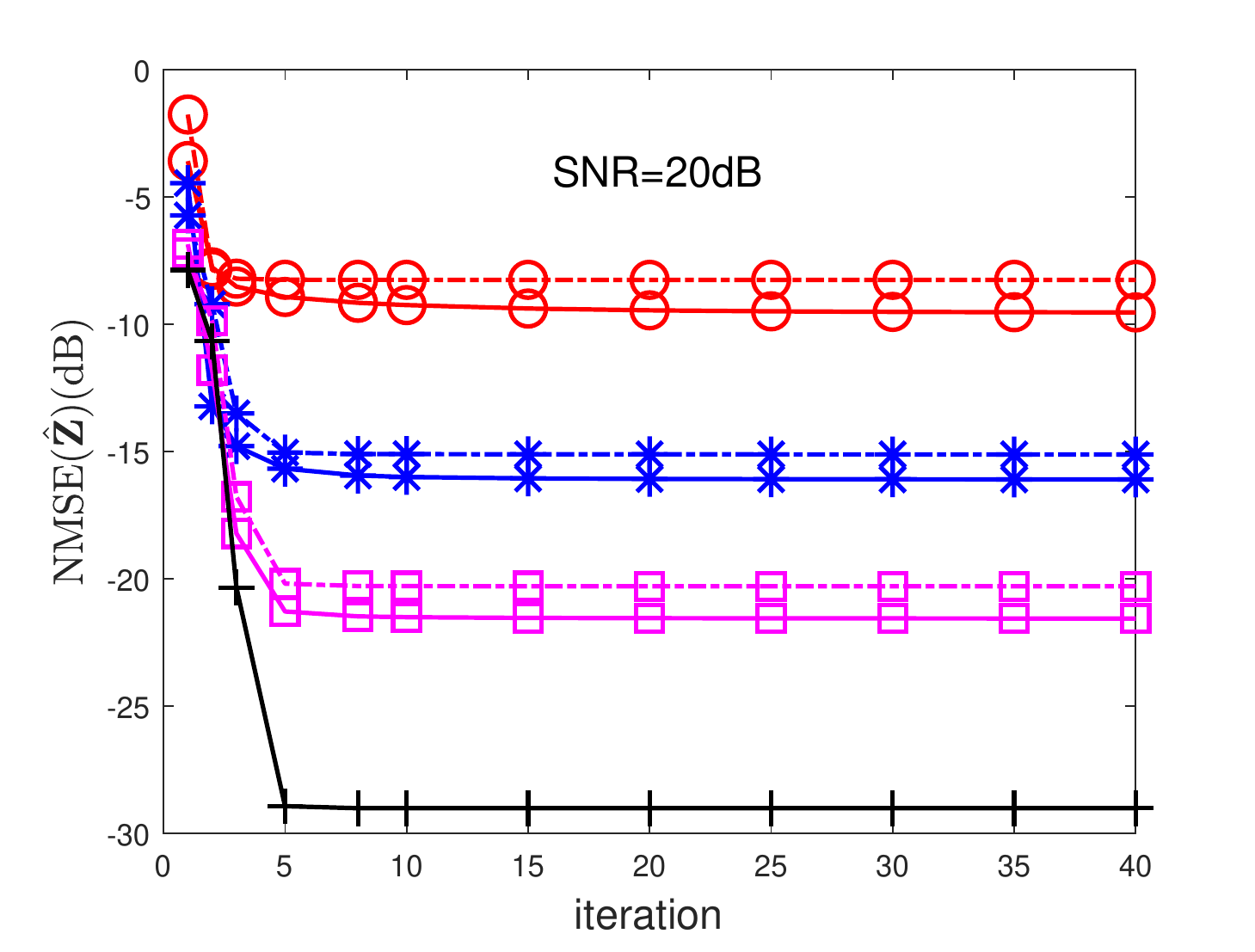}
}
\caption{The NMSE versus iteration for $m=n=100$, $r=5$, $p=0.8$: (a) ${\rm SNR}=0$ dB, (b) ${\rm SNR}=10$ dB, (c) ${\rm SNR}=20$ dB.}\label{NMSEvsIter}
\end{figure}

At first, the NMSE ${\rm NMSE}(\hat{\mathbf Z})$ versus the iteration is investigated and results are shown in Fig. \ref{NMSEvsIter}. It can be seen that MC-Gr-SBL converges in tens of iterations and performs better than VSBL under quantized measurements, especially at high SNR, which demonstrates that taking quantization into account improves the matrix completion performance. In addition, as bit-depth increases, the performances of MC-Gr-SBL improve and approach to VSBL.
\begin{figure}[htbp]
\centering
\subfigure[NMSE versus SNR]{
\includegraphics[width=2.8in]{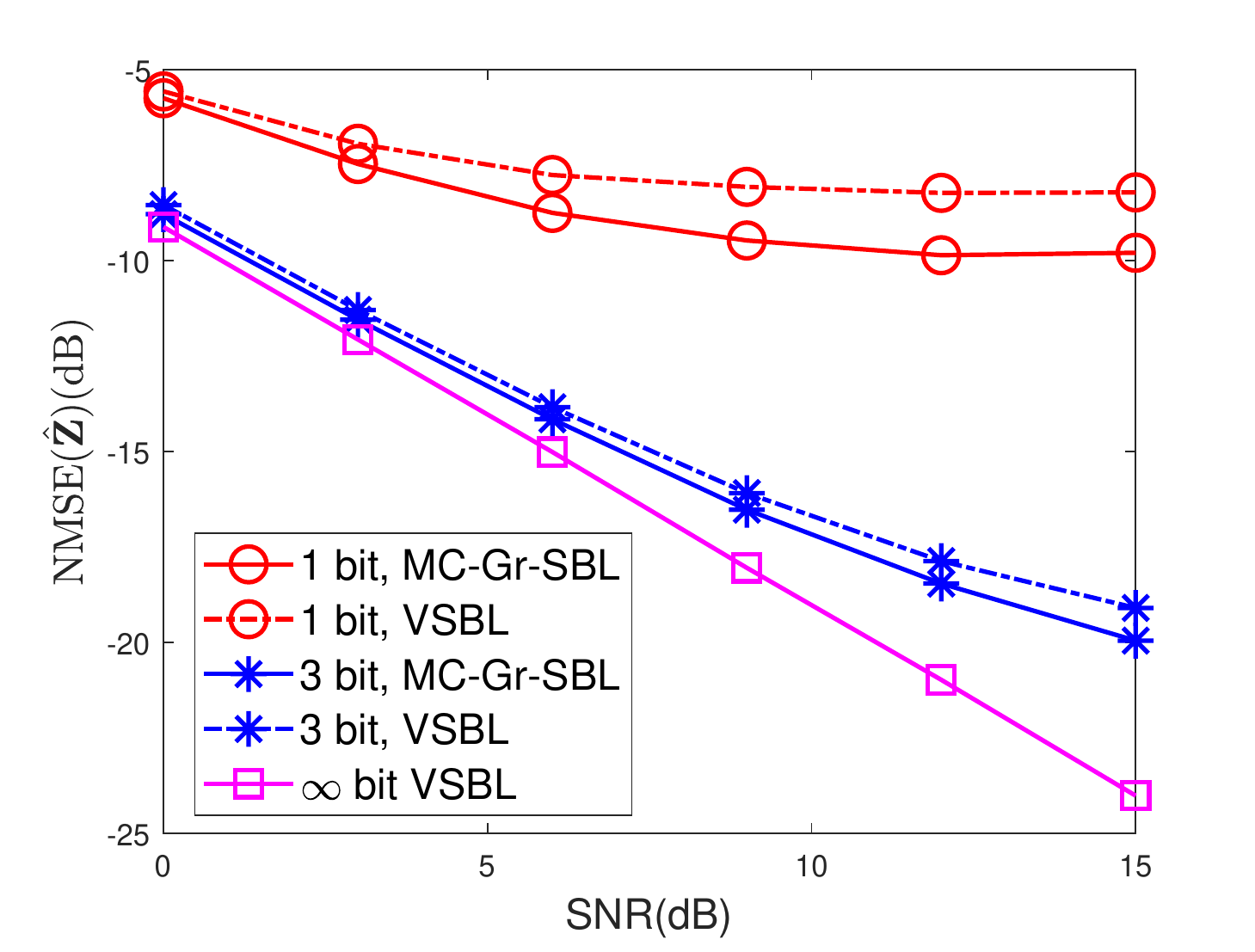}
}
\quad
\subfigure[NMSE versus fraction of entries sampled $p$]{
\includegraphics[width=2.8in]{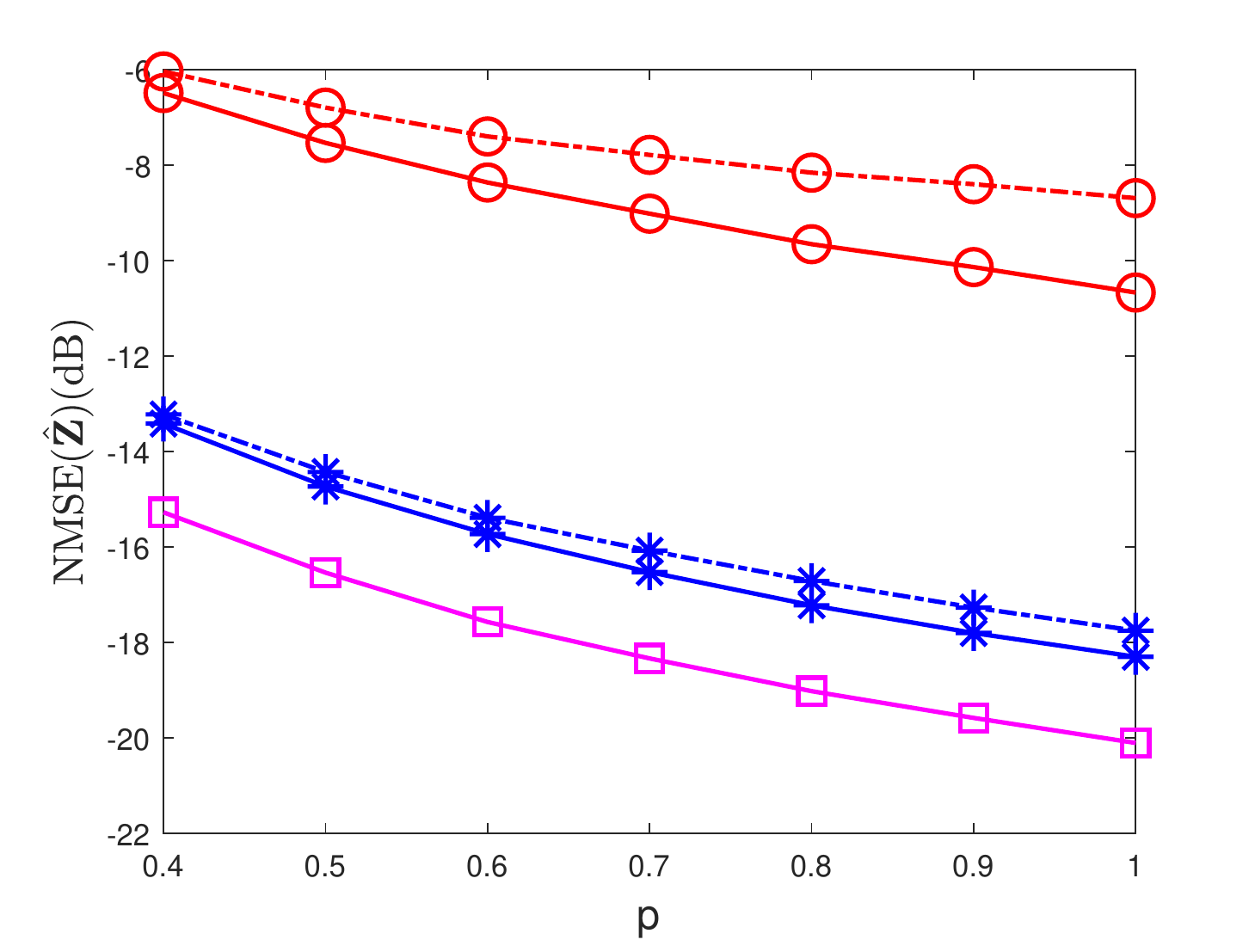}
}
\quad
\subfigure[NMSE versus the number of rows $m$]{
\includegraphics[width=2.8in]{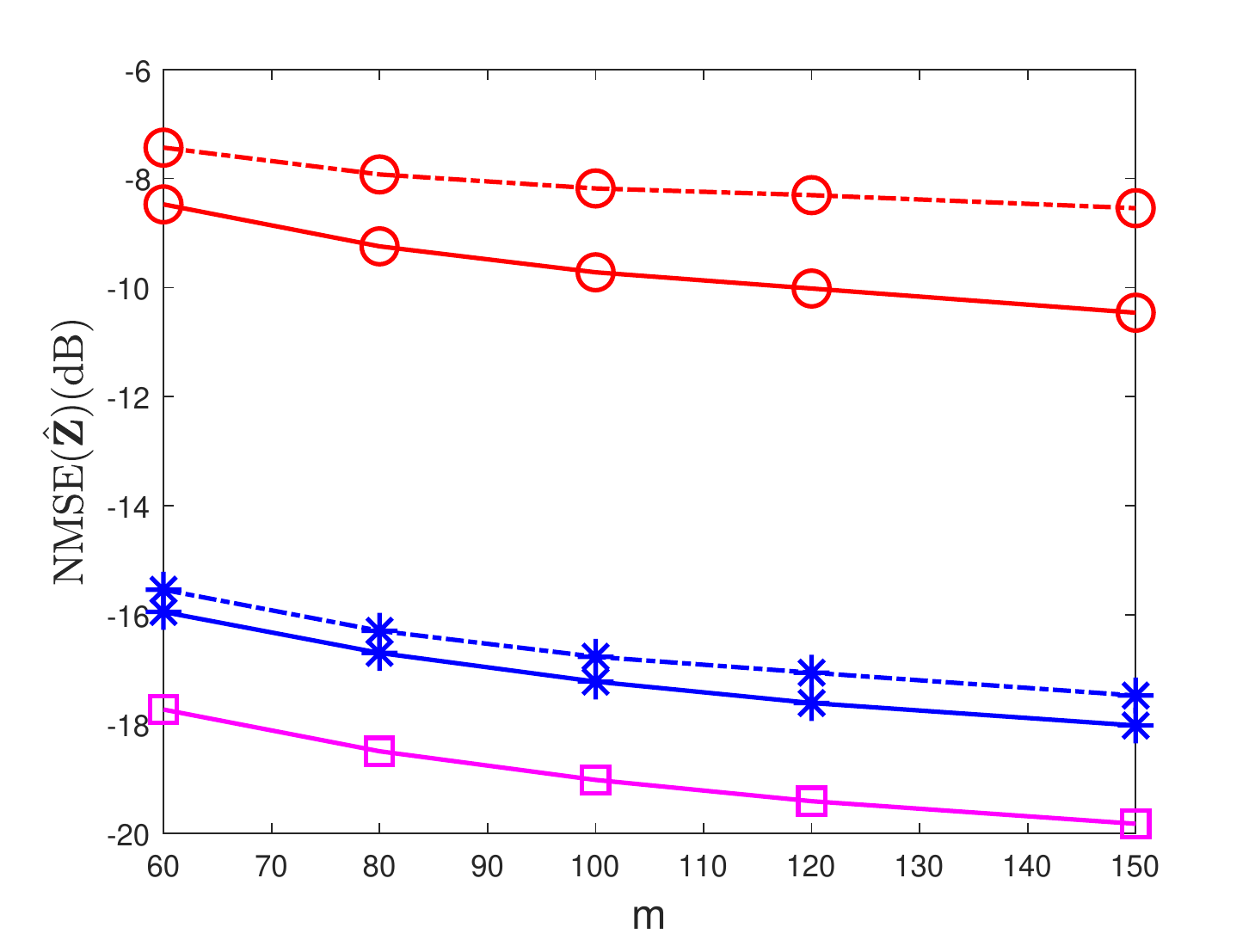}
}
\quad
\subfigure[NMSE versus the rank $r$]{
\includegraphics[width=2.8in]{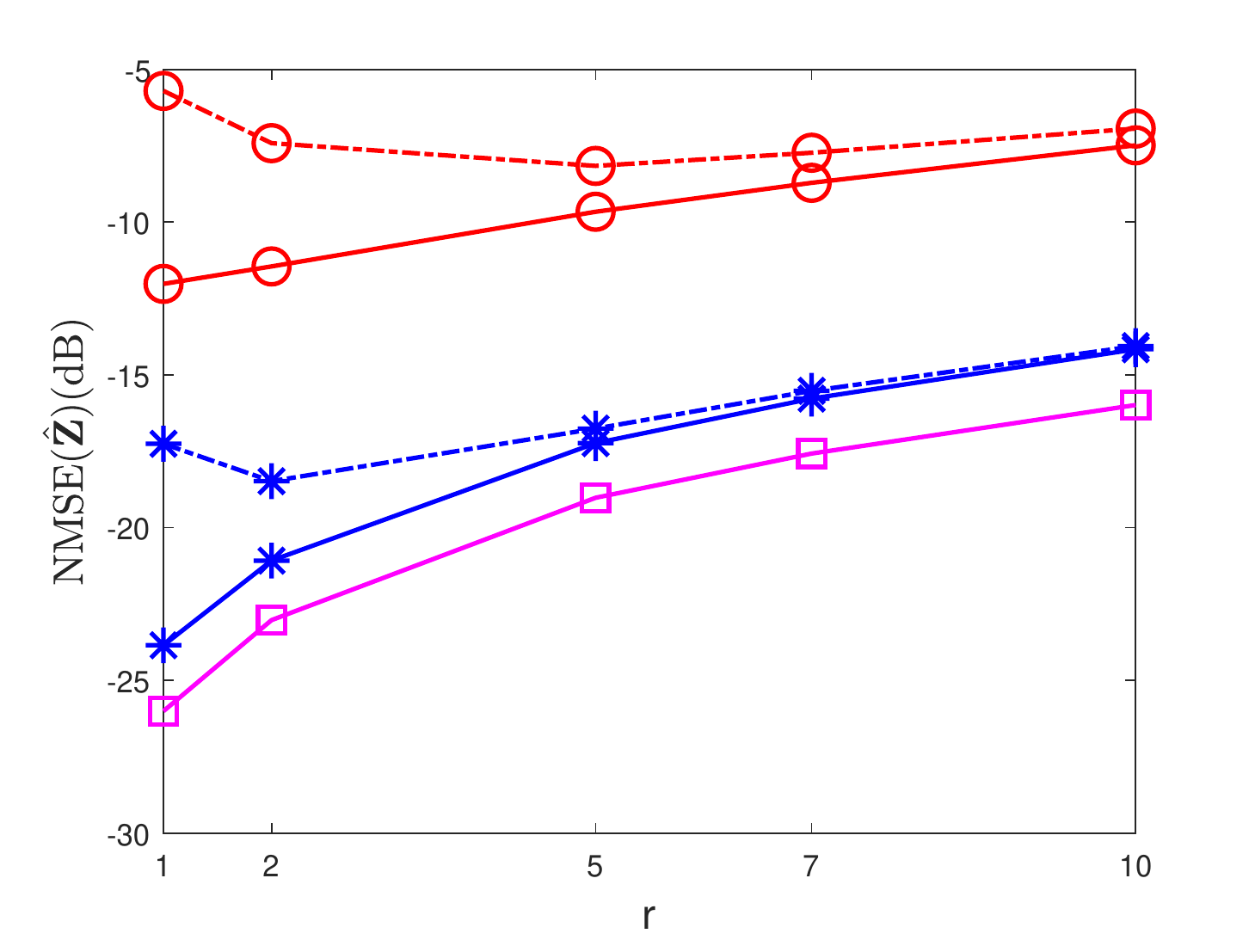}
}
\caption{The NMSE versus SNR, fraction of entries sampled $p$, the number of rows $m$, the rank $r$: (a) Varied SNR and $m=n=100$, $r=5$, $p=0.8$, (b) Varied fraction of entries sampled $p$, $m=n=100$, $r=5$, ${\rm SNR}=10$ dB, (c) Varied number of rows $m$, and $n=100$, $r=5$, $p=0.8$, ${\rm SNR}=10$ dB. (d) Varied rank $r$ and $m=n=100$, $p=0.8$, ${\rm SNR}=10$ dB.}\label{Random_NMSE}
\end{figure}
Then, the NMSE versus SNR, fraction of entries sampled $p$, the number of rows $m$, the rank $r$ are investigated, and results are shown in Fig. \ref{Random_NMSE}. From Fig. \ref{Random_NMSE} (a), it can be seen that as SNR increases from $0$ dB to $15$ dB, the NMSE decreases. For the unquantized setting, the NMSE decreases linearly with respect to SNR. For $1$ bit and $3$ bit quantization, the NMSE decreases quickly when SNR increases from $0$ dB to $6$ dB, then the NMSE decreases slowly when SNR continues increasing. The performance gap between MC-Gr-SBL and VSBL becomes larger as SNR increases.
From Fig. \ref{Random_NMSE} (b)-(d), it can be seen that MC-Gr-SBL performs better than VSBL under quantized setting. Besides, the performance of MC-Gr-SBL improves as the fraction of entries sampled $p$ or the number of rows $m$ increases, or the rank $r$ decreases.




\subsection{Evaluation Under 2D LSE}
\begin{figure}[htbp]
\centering
\subfigure[${\rm NMSE}(\hat{\mathbf Z})$ versus SNR]{
\includegraphics[width=2.8in]{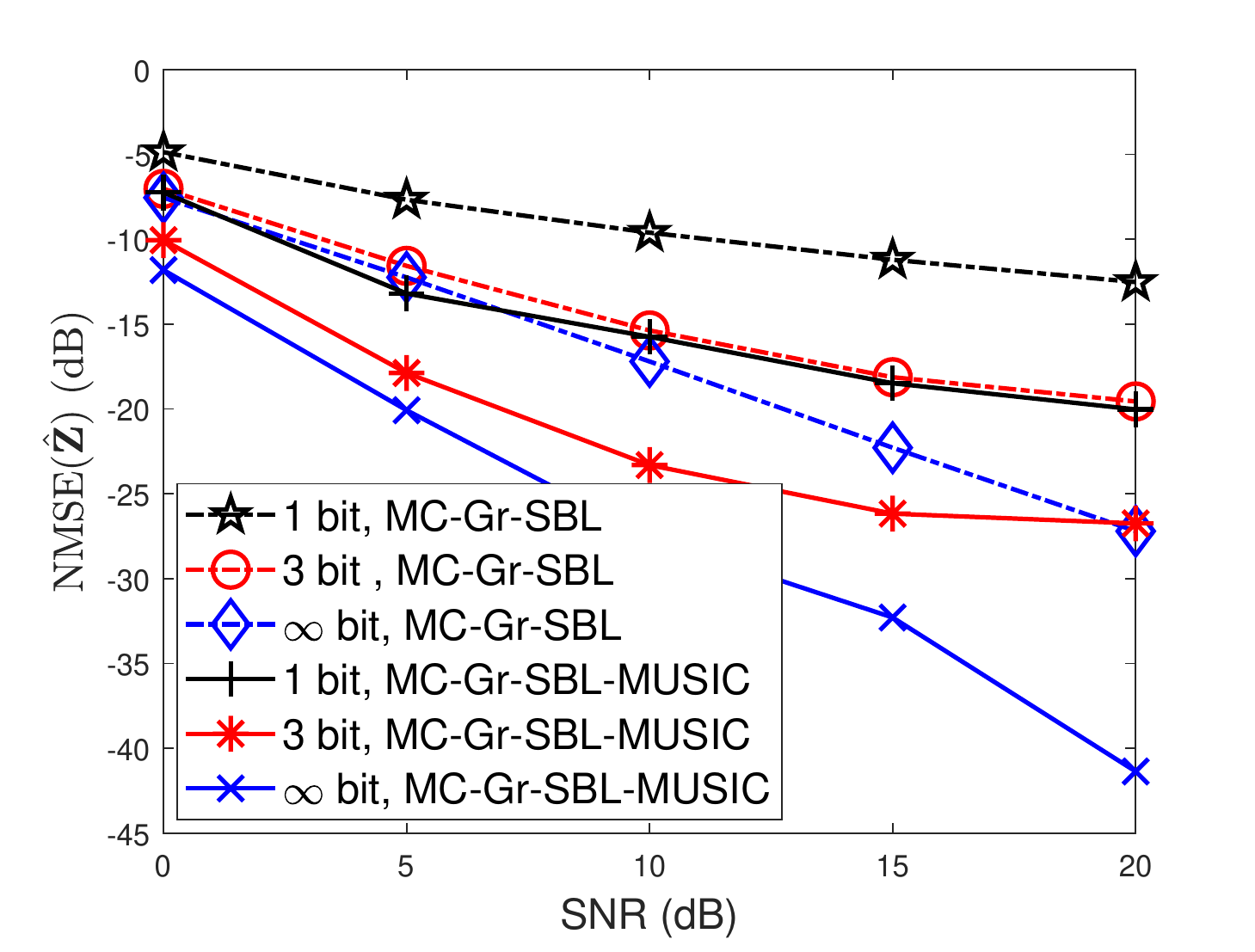}
}
\quad
\subfigure[${\rm P}(\hat{r}=r)$ versus SNR]{
\includegraphics[width=2.8in]{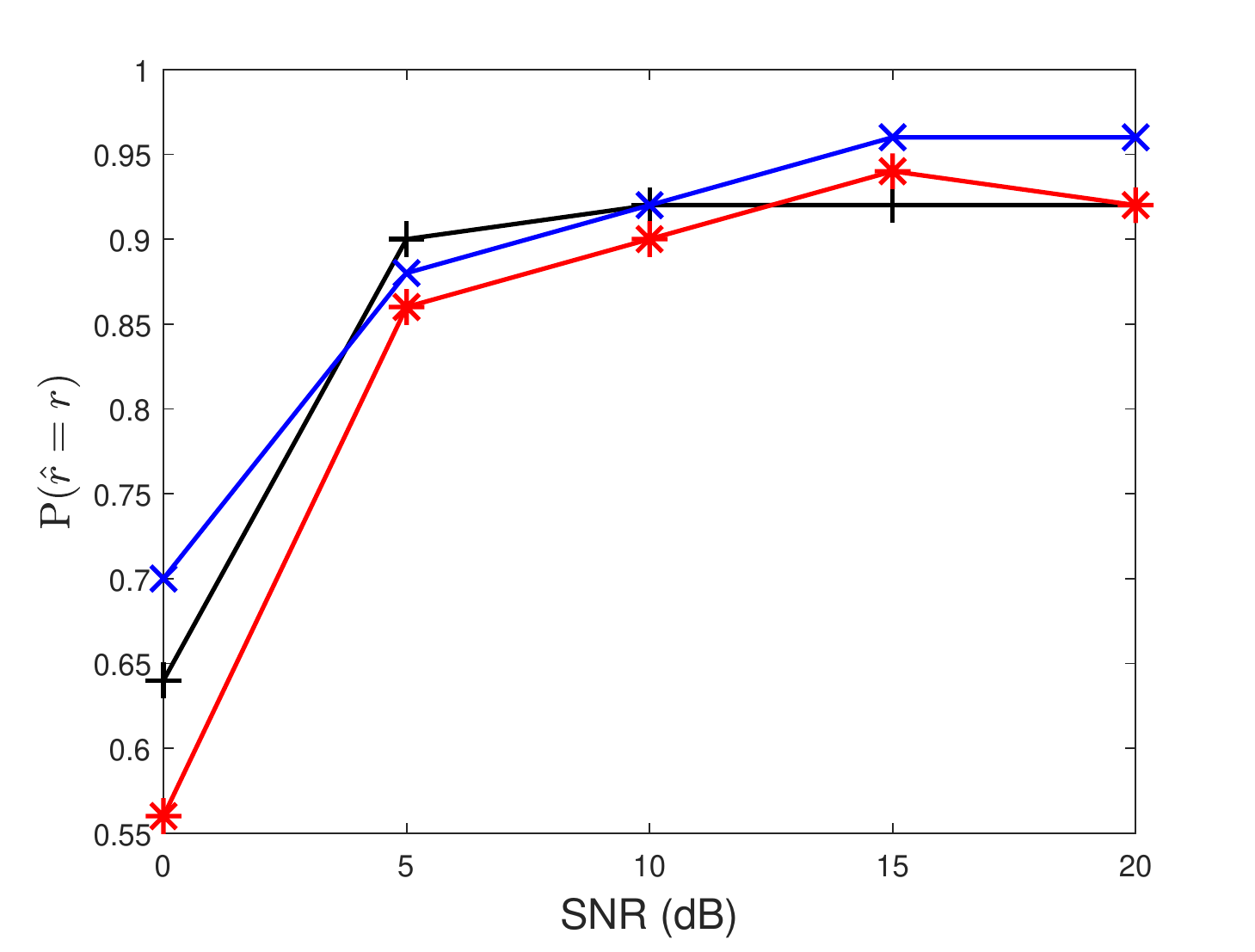}
}
\quad
\subfigure[${\rm MSE}(\hat{\boldsymbol \theta})$ versus SNR]{
\includegraphics[width=2.8in]{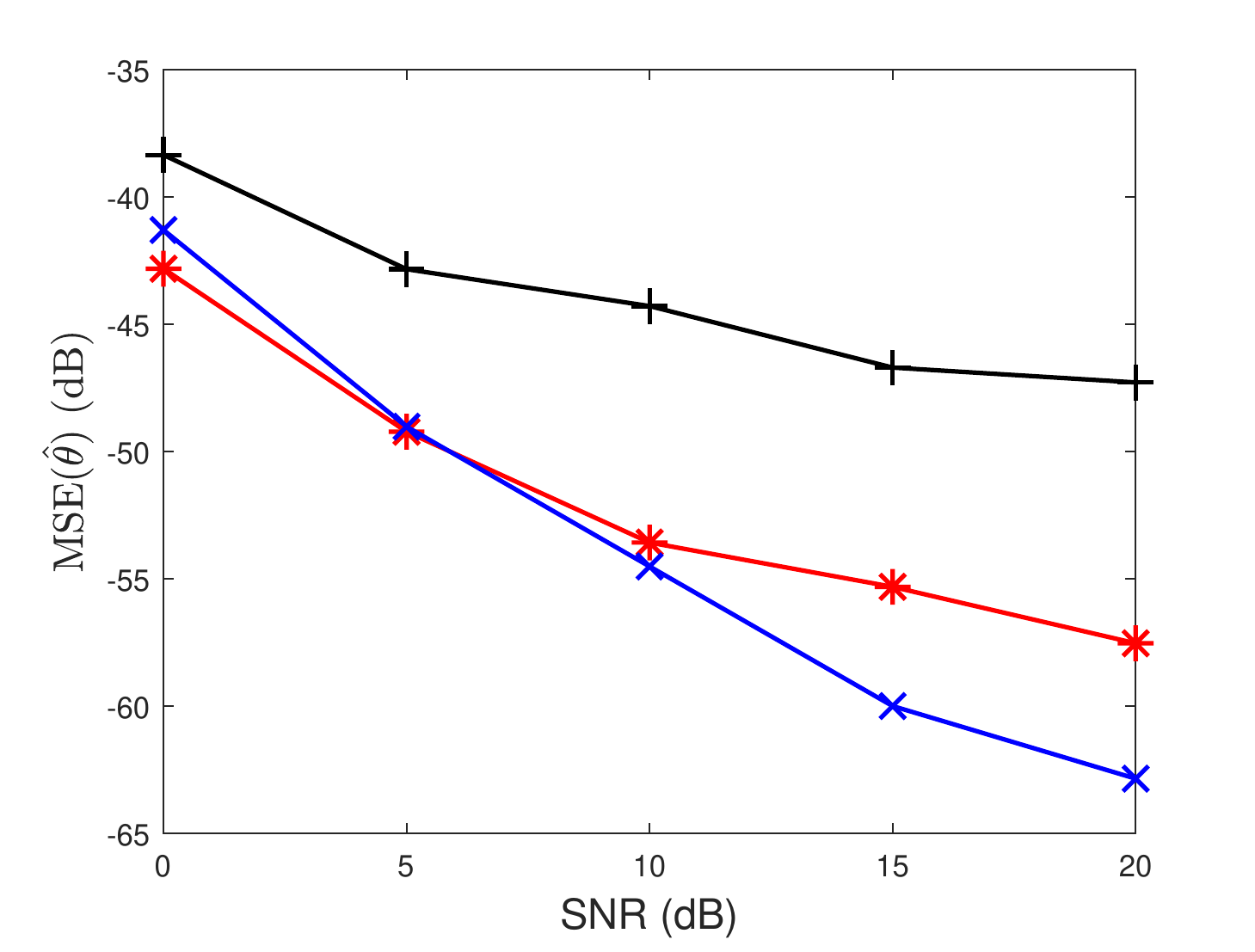}
}
\quad
\subfigure[${\rm MSE}(\hat{\boldsymbol \phi})$ versus SNR]{
\includegraphics[width=2.8in]{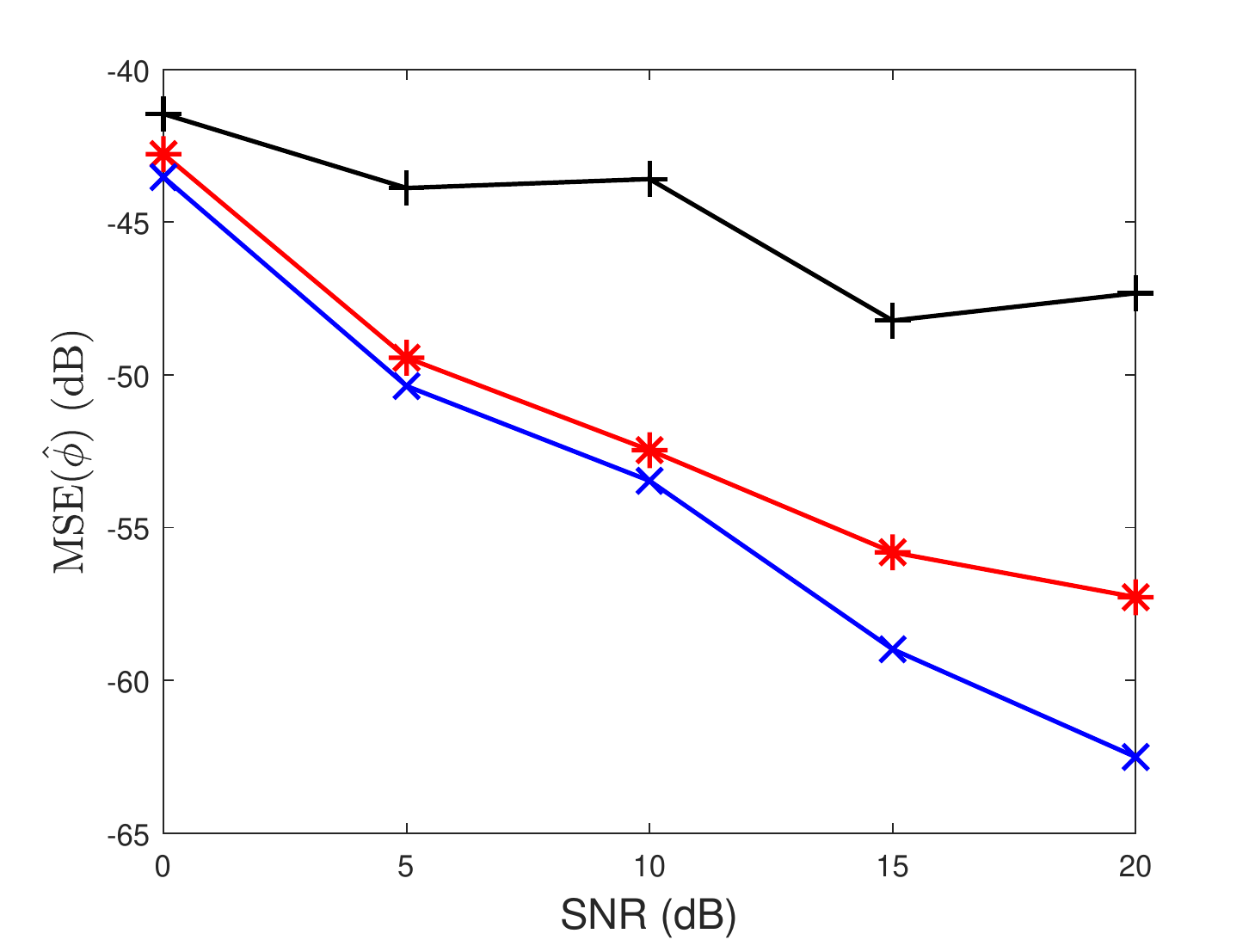}
}
\caption{The 2D LSE results versus SNR for $m=n=40$, $r=3$, $p=0.8$: (a) The NMSE of signal reconstruction error, (b) the correct probability of estimating rank, (c) the frequency estimation error ${\rm MSE}(\hat{\boldsymbol \theta})$, (d) the frequency estimation error ${\rm MSE}(\hat{\boldsymbol \phi})$.}\label{DOA_NMSE}
\end{figure}

The performance of MC-Gr-SBL-MUSIC versus SNR is presented in Fig. \ref{DOA_NMSE}. From Fig. \ref{DOA_NMSE} (a), MC-Gr-SBL-MUSIC performs better than MC-Gr-SBL, which demonstrates that MC-Gr-SBL-MUSIC benefits from utilizing the angular structure. Overall, the probability of successfully estimating the rank improves as SNR increases, as shown in Fig. \ref{DOA_NMSE} (b). As for the frequency estimation error, the MSE of both $\hat{\boldsymbol \theta}$ and $\hat{\boldsymbol \phi}$ achieved by MC-Gr-SBL-MUSIC is very low. For example, at ${\rm SNR}=5$ dB, the MSEs of both $\hat{\boldsymbol \theta}$ and $\hat{\boldsymbol \phi}$ are below $-40$ dB under $1$ bit quantization, demonstrating the effectiveness of MC-Gr-SBL-MUSIC.

\section{Conclusion}
In this paper, MC-Gr-SBL algorithm is proposed to solve the low rank matrix estimation problem from quantized samples. In addition, MC-Gr-SBL combining with MUSIC termed as MC-Gr-SBL-MUSIC is proposed to solve the two dimensional line spectral estimation problem. Numerical results demonstrate the effectiveness of the proposed method.

\end{document}